# Sensing Single Molecules with Carbon-Boron-Nitride Nanotubes

Laith Algharagholy[†a,b,d], Thomas Pope[†c], Qusiy Al-Galiby[†b,e], Hatef Sadeghi[†b], Steve W.D. Bailey[†b], and Colin J. Lambert[†b*]

[a.] College of Computer Science and Mathematics, Al-Qadisiyah University, Diwaniyah, 58002, IRAQ.
[b.] Quantum Technology Centre, Lancaster University, Lancaster LA1 4YB, UK.
[c.] Physics Department, Catania University, 95123 Catania, Italy.
[d.] College of Basic Education, Sumer University, Al-Refayee, 64001, IRAQ
[e.] Physics Department ,Al-Qadisiyah University, Diwaniyah, 58002, IRAQ

**Abstract.** We investigate the molecular sensing properties of carbon nanotube-boron nitride-carbon nanotube (CNT-BN-CNT) junctions. We demonstrate that the electrical conductance of such a junction changes in response to the binding of an analyte molecule to the region of BN. The change in conductance depends on the length of the BN spacer and the position of the analyte and therefore we propose a method of statistically analysing conductance data. We demonstrate the ability to discriminate between analytes, by computing the conductance changes due to three analytes (benzene, thiol-capped oligoyne and a pyridyl-capped oligoyne) binding to junctions with five different lengths of BN spacer.

## 1. Introduction

Chemical sensors that work as electronic noses have attracted extensive attention, because they possess high sensitivity and selectivity towards target analytes, ranging from metal ions and anions to organic neutral chemicals and biological molecules[1-3]. Label-free methods for detecting small molecules are a desirable target technology, because they avoid the need for chemical modification or separation of the analytes, potentially leading to lower costs. One approach to developing such sensors involves measuring the electrical conductance of single-molecule junctions[4-12]. In principle such devices are capable of detecting a single analyte molecule, but controlling their junction separation and stability is difficult[13-15]. Other techniques for molecular sensing[16] involve measuring the change in electrical conductance of carbon nanotubes in response to molecular[17] or changes in their vibrational[18]. In this paper, our aim is to build upon such approaches by demonstrating that single-molecule sensing capabilities can be significantly improved by utilising carbon/boron-nitride/carbon hetero nanotube junctions. Such nanotubes can be regarded as sculpturenes; ie novel nanometre-scale objects, obtained by cutting selected shapes from layered materials and allowing the shapes to reconstruct[19-21]. The simplest examples of sculpturenes are formed by cutting straight nanoribbons from bilayer graphene and allowing the edges to reconstruct to maximise $sp^2$ bonding. If the width of the nanoribbon is sufficiently small (i.e. of order 3nm or less) then the whole ribbon can reconstruct to form a carbon nanotube (CNT), with a pre-defined location and chirality. This cutting can be achieved using lithographic[22-24], chemical[25-28] or sonochemical[29, 30] techniques. If the graphene layers are contacted with a boron-nitride (BN)[31], the reconstructed nanotube will be of a hetero structure. Previous studies have shown that the electronic properties and the stability of such hetero structures depend on the configuration of the B, N and the C atoms. It has been shown that doping an armchair CNT with a BN region leads to a tuneable HOMO-LUMO gap[32, 33]. It is also known that the CNT-BN interface leads to localized states[20, 34-36]. These states are present in doped nanotubes[37-40] and similar interfaces in graphene [31].

In what follows we investigate the sensing capabilities of a CNT-BN-CNT structure formed from two (6,6) CNTs connected via an equivalent (6,6) BN nanotube. An electrical current flows through the BN from one CNT to the other and our aim is to understand the change in conductance of such a structure when a single analyte molecule binds to it. Since the BN possesses a large energy gap around the Fermi energy[38, 41-43], it behaves as an insulating barrier, which fixes the distance between the two CNT electrodes. We shall demonstrate that when an analyte binds to the BN, the change in conductance depends on both the nature of the analyte and on the fixed length of the BN barrier layer. By analysing the response of devices with different BN-lengths, a unique fingerprint to each analyte is acquired which can be used for discrimination.

## 2. Characterizing the junction

Figure 1 shows five sculpturene junctions (labelled a-e) constructed from two (6, 6) armchair CNTs connected by BN barrier layers of lengths ranging from one to five unit cells. To construct these junctions, we relaxed hetero nanoribbons using the SIESTA implementation of DFT [44] to minimise the forces on the atoms. In all cases, we used the Ceperley-Alder (CA) exchange correlation functional, with norm-conserving pseudopotentials and double zeta polarized (DZP) basis sets of pseudo atomic orbitals.

The transmission coefficients, T(E), for electrons of energy E passing through the BN barrier are obtained using the Green's function-based transport code GOLLUM[45], which utilises the DFT-based hamiltonian from SIESTA. We show in figure 2 the transmission coefficients for each junction in isolation. When E lies within the band gap of the BN, T(E) decays

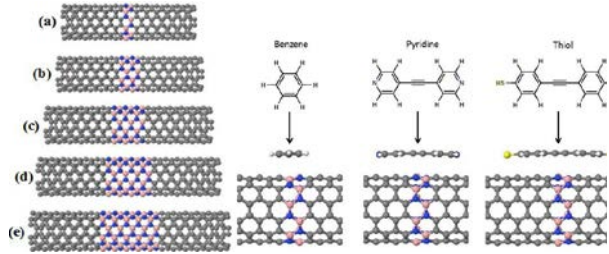

**Fig 1:** (Left) Hetero junctions constructed from two (6,6) Carbon nanotubes joined by a BN insulating layer of varying length. (Right) an example of CNT-BN with three anylate molecules ( benzene, pyridine and thiol).

exponentially with the length of the BN barrier as

$$T(E) = T_c \, e^{-\beta x} \quad (1)$$

where $T_c$ represents the effects of scattering at the BN-C interface and $x$ is the barrier length. At $E=E_F$, where $E_F$ is the Fermi energy (ie Dirac point) of the CNTs, we calculate that the attenuation factor, $\beta = -3.6\pm0.2$ (see figure 3), which is of the same order as that of a typical oligoyne [46].

The local density of states at the Fermi energy for the 1BN hetero junction was computed in Fig S1 of SI, a state associated with the N-C interface can be seen and since this state bridges the junction, it is responsible for a small peaks in T(E) near $E_F$. This can be seen most prominently in the transport curves (Fig S2 of SI) for the 2BN and 3BN junctions. Since the transport through the 1BN junction is high irrespective of this effect, the peaks are not as clear. The only states near the Fermi energy localised on the BN buffer are these B-C and N-C interface states. Therefore, for a clean junction, they represent the smallest distance between the two electrodes and any effect a molecule has on the junction will depend strongly on the molecule's interaction with these states.

## 3. Discriminating single-molecule sensing of the device

To test the sensing capability of each device, we placed analyte molecules at various locations in the vicinity of the BN spacer and computed the resulting transmission coefficient. Figure 2 shows results for the thiol-capped oligoyne. The dark lines show T(E) for the clean junctions and the families of lighter lines show the T(E) for various analyte binding locations. Results for two other analytes (pyridine-capped oligoyne and benzene) are shown in figures S5 and S6.

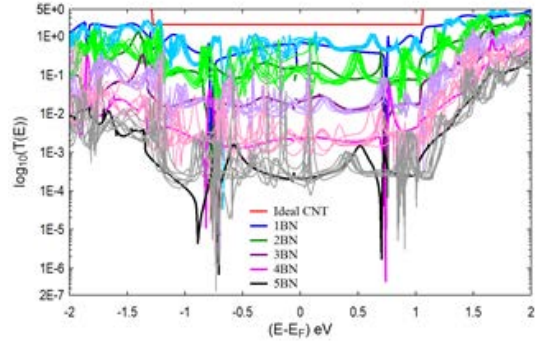

**Fig 2:** Plot of the logarithm of the transport of the thiol-capped oligoyne-doped junction, $log_{10}(T(E))$, for 1BN to 5BN. The darker lines represent the corresponding transport of the clean junction – also shown in Fig 1. For each junction, the thiol-capped oligoyne was placed at several locations.

These figures show that the transmission coefficient fluctuates with different binding locations and therefore to sense and discriminate the analytes, a statistical approach to data analysis is needed.

In what follows, we define $T_{X,m}(E)$ to be the transmission coefficient in the presence of analyte $X$, (where $X$ =benzene, a pyridine-capped oligoyne (PY) or a thiol-capped oligoyne (SH)) located at position $m$. From the Landauer formula, the corresponding electrical conductance is $G_{X,m}(E_F) = G_0\, T_{X,m}(E_F)$. Therefore we define the quantity $\alpha_{X,m}(E_F)$ which is a measure of the difference between $G_{X,m}(E_F)$ and the conductance $G_{bare}(E_F)$ in the absence of a dopant [47, 48]:

$$\alpha_{X,m}(E_F) = log_{10}G_{X,m}(E_F) - log_{10}G_{bare}(E_F) \qquad (2)$$

To differentiate between different junctions, we analyse the set of all values of $\alpha_{X,m}(E_F)$ for $E_{min} < E_F < E_{max}$ and configuration $m = 1,…, M$ belonging to a given analyte $X$. These values can be obtained experimentally by using a third gate electrode to sweep through a range of Fermi energies. The probability distribution of the set $\{\alpha_{X,m}(E_F)\}$ for a given $X$ is then defined by:

$$P_X(\alpha) = \frac{1}{M(E_{max} - E_{min})} \sum_{n=1}^{M} \int_{E_{min}}^{E_{max}} dE\, \delta(\alpha - \alpha_{X,m}(E)) \qquad (3)$$

where $\delta(\alpha - \alpha_{X,m}(E))$ is a Dirac delta function.

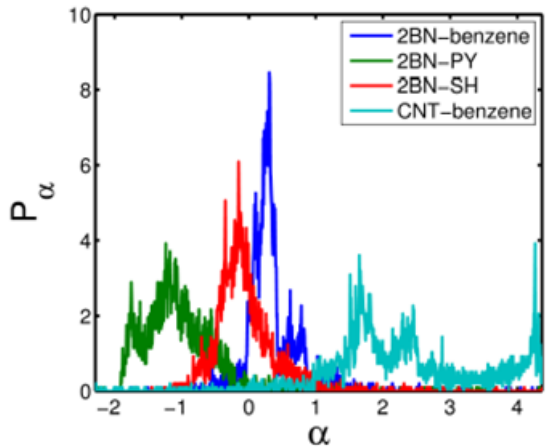

**Fig 3:** shows probability distribution $P_X(\alpha)$ of the set $\{\alpha_{X,m}(E)\}$ for a 2BN junction with benzene (blue), 2BN junction with PY (green), 2BN junction with SH (red) and ideal CNT with benzene (turquoise).

To demonstrate that a CNT-BN-CNT heterostructure is a more discriminating single-molecule sensor than a pure CNT, the magenta curve in Fig 3 shows the probability distribution $P_x(\alpha)$ for benzene adsorbed on a pure CNT while the blue curve shows the corresponding distribution for a hetero-tube containing a 2BN barrier. Clearly the latter is a more narrow distribution, which facilitates discriminating sensing. To demonstrate discrimination, the green and red curves in figure 3 show the distributions arising when Py-capped oligoynes and SH- capped oligoynes bind to the 2CN hetero nanotube. These results for the three different analytes were obtained by sampling the curves at a uniformly spaced set of energies and creating histograms of the associated values of $\alpha_{X,m}(E)$. The calculations of the probability distributions for other hetero-nanotubes with different lengths of BN spacer are shown in Fig S7 of the SI.

### 4. Conclusions

Figure 3 shows that the different analytes lead to peaks in $P_x(\alpha)$ located at different values of $\alpha$ and therefore the 2BN structure is able to discriminate. This is in marked contrast with a pure CNT, which possesses rather broad peaks consequently a reduced the ability to discriminate. This broadening is illustrated in figure 3 which shows $P_x(\alpha)$ for benzene adsorbed on a (6,6) CNT with no barrier. As well as showing a high sensitivity to the presence of a molecule, the junctions also show promising results suggesting a high level of selectivity. Since these junctions are reusable, they offer great versatility in probing the electronic properties of analytes. More importantly, the junctions are stable. In practice, the atomic-scale detail of hetero-junctions will not be known and will vary from sample to sample. Therefore it will be necessary to calibrate each junction for the range of analytes of interest and to preserve the calibration, the junctions must be stable. This feature of CNT-BN-CNT junctions makes them particularly attractive compared with single-molecule junctions, which usually are not stable over long periods of time.

### Acknowledgments
This work is supported by the UK EPSRC, EP/K001507/1, EP/J014753/1, EP/H035818/1, and from the EU ITN MOLESCO 606728 and the Ministry of Higher Education and Scientific Research, Al Qadisiyah University and Thi-Qar University, IRAQ.